\documentclass{PoS}
\usepackage{psfrag}

\newcommand{\be}{\begin{equation}}
\newcommand{\ee}{\end{equation}}
\newcommand{\bea}{\begin{eqnarray}}
\newcommand{\eea}{\end{eqnarray}}
\newcommand{\bi}{\begin{itemize}}
\newcommand{\ei}{\end{itemize}}

\newcommand{\Dw}{D_{\rm w}}

\newcommand{\DWF}{D_{\rm DWF}}
\newcommand{\Deff}{D_{\rm eff}}

\newcommand{\xbf}{{\bf x}}
\newcommand{\ybf}{{\bf y}}

\title{A formulation of domain wall fermions in the Schrodinger functional}

\ShortTitle{A formulation of domain wall fermions in the Schrodinger functional}

\author{
        \speaker{Shinji Takeda}\\
       Department of Physics, Columbia University\\
       New York, NY 10027, USA\\
       E-mail: \email{st2542@columbia.edu}}

\abstract{
We present a formulation of domain wall fermions in the Schroedinger functional
by following the universality argument given by L\"uscher.
To check whether the formulation works,
we examine the lowest eigenmode of the free domain wall fermion operator.
We confirm that the theory belongs to a correct universality class
and that the eigenvector is localized near the
boundaries of the fifth dimension.
We also investigate the chiral symmetry breaking
structure of the four dimentional effective operator.
We observe that the bulk chiral symmetry breaking
disappears for a large fifth dimensional size,
while the breaking originated by the boundary effects persists
and exponetially decays away from the time boundaries.
}

\FullConference{The XXVII International Symposium on Lattice Field Theory\\
		 July 26-31, 2009\\
		 Peking University, Beijing, China}

\begin{document}

\section{Introduction}

In the study of CP violation by CKM unitary triangle analysis,
hadron matrix elements of four-fermion operators,
such as $B_{\rm K}$, play a vital role.
Accurate calculations of this quantity from first principles
are an important task for the lattice QCD community.
In such calculations, having chiral symmetry 
is crucial to avoid an operator mixing problem
which causes uncontrollable systematic errors.
Therefore, the RBC/UKQCD collaboration \cite{Allton:2008pn} is currently using
lattice chiral fermions, such as, domain wall fermions (DWF),
to compute $B_{\rm K}$.
In the course of the computation,
there are many sources of systematic errors which one has to control.
Among them, the non-perturbative renormalization (NPR)
could be a serious source.
At the moment, the collaboration has been using conventional schemes, namely,
the RI/MOM scheme and its variant \cite{Aoki:2007xm,Sturm:2009kb}.
However, this scheme could be subtle
depending on the target quantity.
To avoid such difficulties,
a new scheme was invented, known as the Schrodinger functional
(SF) scheme \cite{Luscher:1992an}.
This scheme provides a reliable way of estimating errors in the NPR.
If one wants to use this scheme for the renormalization
of $B_K$ given by the RBC collaboration,
first of all, one has to formulate DWF in the SF setup.
This is a purpose of this paper.

In the SF setup,
a formulation of lattice chiral fermions
is in fact a non-trivial task.
This can be seen as follows.
First, let us consider the continuum theory
with the SF boundary conditions,
\bea
P_+\psi(x)|_{x_0=0}
&=&
P_-\psi(x)|_{x_0=T}=0,
\\
\bar\psi(x)P_-|_{x_0=0}
&=&
\bar\psi(x)P_+|_{x_0=T}=0,
\eea
with $P_\pm=(1\pm\gamma_0)/2$.
These boundary conditions break chiral symmetry explicitly.
This fact indicates that the fermion propagator
in such a theory does not anti-commute with $\gamma_5$ and
breaking term \cite{Luscher:2006df} is given by
\be
\gamma_5 D^{-1}+D^{-1}\gamma_5=\Delta.
\label{eqn:continuum}
\ee
The breaking term
$\Delta$ is supported at the time boundaries,
whereas the anti-commutation relation holds in a bulk.
This is a situation of the continuum theory.
On the other hand, if one naively defines lattice chiral
fermions by using the SF Wilson kernel operator,
such a chiral operator automatically satisfies the Ginsparg-Wilson (GW) relation
\cite{Ginsparg:1982bj}
and its inverse cannot reproduce the continuum results in
eq.(\ref{eqn:continuum}).
This shows a difficulty of formulating the chiral fermions on the SF.

To overcome such a situation,
Taniguchi \cite{Taniguchi:2004gf} made the first attempt of
formulating overlap fermions in the SF
by using an orbifolding technique.
Furthermore, he extended his idea to DWF \cite{Taniguchi:2006qw}
and then he and his collaborators obtained the renormalized
$B_K$ in the quenched QCD \cite{Nakamura:2008xz}.
However, in dynamical simulations,
his approach for overlap fermions has a sign problem for odd flavors.
Although this problem can be removed for even flavors,
the flavor symmetry is broken explicitly as an expense.
Sint \cite{Sint:2005qz} proposed
a solution for even numbers of flavors which
preserves the flavor symmetry exactly.
However, this formulation can only be applied
for the even number of flavors.
This conflicts with the current trend
toward large scale
dynamical $2+1$ flavor simulations.
To overcome such a circumstance,
L\"uscher \cite{Luscher:2006df} proposed a completely different approach
following the universality and symmetry considerations.
In that paper, he gave a formulation for overlap fermions.
However, DWF from this approach has not been discussed so far,
therefore we address this issue in this paper.

In the following, after introducing L\"uscher's argument,
we give a formulation of DWF in the SF setup.
And then we show some investigations
to check how it really works.
To enhance readability, in the following,
the lattice spacing is suppressed unless necessary.

\section{Universality argument}
The argument in the continuum theory given in the previous section
suggests that
on the lattice, the GW relation has to be modified to
correctly reproduce the continuum results.
To this end,
a chiral fermion operator also has to be modified compared with
that of the usual infinite lattice.
It is not so hard to break the GW relation itself
and one can easily imagine that such a modification
would be local and related with the boundary conditions.
However, it is not clear how the SF boundary conditions come out?
L\"uscher gave an answer to the question \cite{Luscher:2006df}, which uses a
symmetry consideration and an order counting.
What he concludes is that the SF boundary conditions are natural and
more stable than other boundary conditions in the continuum limit,
because the dimension of the Dirichlet type boundary conditions
is lower than that of the Neumann type boundary conditions.
Bottom line is that
it is guaranteed that the SF boundary conditions are automatically
reproduced in the continuum limit without fine tuning
if the lattice fermion operator is modified such that the GW relation
is broken near the time boundaries.

Before going into the concrete discussion of DWF,
let us summarize a general instruction to formulate chiral fermions in the SF
from the L\"uscher's argument.
Firstly, including additional terms, which break the GW relation,
to a lattice fermion operator such that
the GW breaking is supported near the time boundaries.
Secondly, keeping other symmetries,
like C, P, T and flavor symmetries etc., which we do not want to lose.
This instruction does not dictate a unique form of the operator,
and actually there are an infinite number of possibilities.
The important thing here is that
the details of the operator actually do not matter
as long as it fulfills these conditions.
When these conditions hold, the corresponding lattice
fermions are automatically guaranteed to follow
the SF boundary condition
in the continuum limit thanks to the order counting argument.
By following this instruction,
we define DWF in the SF in the next section.

\section{Universality formulation of DWF}
A possible form of the massless DWF operator in the SF
is given by
\be
\DWF
=
\left[
\begin{array}{cccccc}
\Dw+1&-P_L&&&&cB\\
-P_R&\Dw+1&-P_L&&cB&\\
&-P_R&\Dw+1&-P_L+cB&&\\
&&-P_R-cB&\Dw+1&-P_L&\\
&-cB&&-P_R&\Dw+1&-P_L\\
-cB&&&&-P_R&\Dw+1\\
\end{array}
\right].
\ee
This is a five dimensional block form and
the block elements are four dimensional operator.
$\Dw(m_5)$ is the Wilson fermion operator in the SF \cite{Sint:1993un}
with the parameter $m_5$ and is supported in a range of the time
$1\le x_0 \le T-1$.
Here, the size for the fifth direction\footnote{We assume that $L_s$ is an even number.}
is taken as $L_s=6$ as an example.
The additional terms which are proportional to the coefficient $c$
are distributed along the cross diagonal elements.
This is something like a generalized mass term
and breaks the chiral symmetry explicitly.
The block element $B$
is now supported only near time boundary,
\be
B(x,y)
=
\delta_{\xbf,\ybf}
\delta_{x_0,y_0}\gamma_5(\delta_{x_0,1}P_-+\delta_{x_0,T-1}P_+).
\label{eqn:boundaryterm}
\ee
This time dependence comes from a motivation that
we do want to maintain the chiral symmetry in a bulk in time.
We will check that the breaking of the GW relation
is really supported near the time boundaries exponentially
(see Section \ref{sec:GWbreaking}).
In this way, such fifth dimensional coordinate and time
dependences are determined.
Next, how about the spinor structure?
Actually, this structure may be fixed to some extent by
requiring the C, P, T symmetries and the $\Gamma_5$-Hermiticity.
These requirements are not so strong to determine
the spinor structure completely and there is some freedom.
The structure given in eq.(\ref{eqn:boundaryterm})
is one of many solutions.
Actually, we examined several choices
of the spinor structure in the boundary term
and confirmed the universal results
in the continuum limit for the lowest eigenvalue.

The coefficient\footnote{$c=0$ is not allowed as we will see in Section \ref{sec:eigen}.}
$c\ne 0$ is a parameter
which can be used to achieve the O($a$) improvement.
The physical mass term can be added
in the same way as the infinite lattice case.

\section{Check of formulation}
The next step is to check how it works.
We investigate several things as follows.
First, we study the lowest eigenmode for the free operator
to see the importance of the presence of the boundary term.
Second, we address the breaking of the GW relation
for the four dimensional effective operator.
This is to see what the chiral breaking term looks like.
We also checked scaling properties of the
free spectrum of the lowest ten eigenvalues
and carried out the one-loop
calculation and obtained sensible results.
However we do not show them here
due to limited space.
In the near future, we will show them in a full paper.

\subsection{The lowest eigenmode for free massless operator}
\label{sec:eigen}

\begin{figure}[t]
\begin{center}
  \begin{tabular}{cc}
  \scalebox{1.6}{\includegraphics{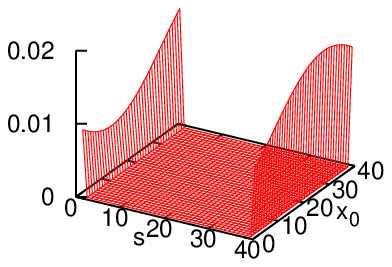}}&
  \scalebox{1.6}{\includegraphics{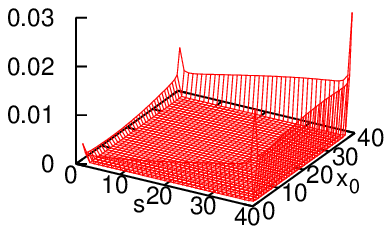}}\\
  \end{tabular}
\caption{$||\psi(x_0,s)||_{\rm spin}$
with the zero spatial momentum with the parameters
$T=L_s=40$, $\theta=0$ and $m_5=1$.
The norm for the eigenvector is taken for the spinor space.
The left (right) panel is for $c=1$ ($c=0$).
\label{fig:WF}}
\end{center}
\end{figure}

Let us see the lowest eigenmode of the free massless operator
$\DWF^{\dag} \DWF$
\be
\DWF^{\dag} \DWF \psi=\lambda_{\min}\psi.
\ee
To show the importance of the boundary term,
we consider two cases with $c=1$ and $c=0$.
In the free case,
the Fourier transformation can be performed
for the spatial directions
and we project onto the zero momentum configuration.

For $c=1$, we obtained the minimum eigenvalue,
$T^2\lambda_{\min}=2.59..$.
This value is rather close to the continuum one
$\pi^2/4=2.47..$ in Ref.\cite{Sint:1995ch}.
The difference is considered to be a lattice artifact.
The eigenfunction in Figure \ref{fig:WF} (left panel)
shows the expected behavior,
namely, being localized in the fifth direction while propagating
in the time direction. This is apparently a physical mode.
On the other hand, when switching off the boundary term $c=0$,
the eigenvalue $0.61...$ is quite far from the continuum value.
Actually, a scaling study with larger lattice sizes
shows that the theory does not belong to
the correct universality class.
Furthermore, the eigenfunction in the right panel
of Figure \ref{fig:WF} appears to be
propagate in the fifth direction.
This is not a physical mode anymore.

This result shows that the presence of the boundary
term is essential to correctly produce the continuum results.
In other words, this term is essential to be
in a correct universality class.

\subsection{Chiral symmetry breaking by boundary effect}
\label{sec:GWbreaking}

Next, let us check that DWF in the SF
has an expected chiral symmetry breaking structure.
To this end, for the effective four dimensional operator
\cite{Kikukawa:1999sy,Kikukawa:1999dk},
\be
\det [\Deff^{(L_s)}]=\det [\DWF/D_{\rm{PV}}],
\ee
let us consider the breaking of the GW relation
\be
\Delta^{(L_s)}
=
\gamma_5\Deff^{(L_s)}+\Deff^{(L_s)}\gamma_5
-2\Deff^{(L_s)}\gamma_5\Deff^{(L_s)}.
\label{eqn:GWbreaking}
\ee
For finite $L_s$, there are two sources of chiral symmetry breaking,
namely, a bulk source and a boundary source.
The former can be removed
by taking $L_s$ to infinity.
After this limit, the remaining chiral symmetry breaking must be due to the
boundary source only.
Actually, it is known that the corresponding breaking for overlap fermions
exponentially decays away from the time boundaries \cite{Luscher:2006df}.
We want to check this kind of phenomena for DWF
in a free case numerically.

\begin{figure}[t]
\begin{center}
  \begin{tabular}{ccc}
  \hspace{-30mm}
  \scalebox{1.2}{\includegraphics{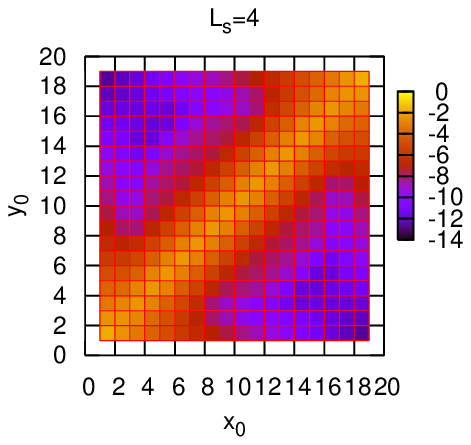}}&
  \hspace{-40mm}
  \scalebox{1.2}{\includegraphics{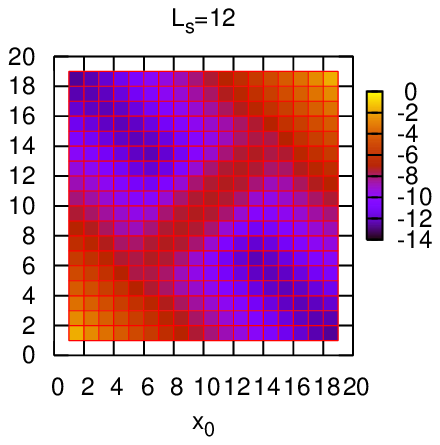}}&
  \hspace{-40mm}
  \scalebox{1.2}{\includegraphics{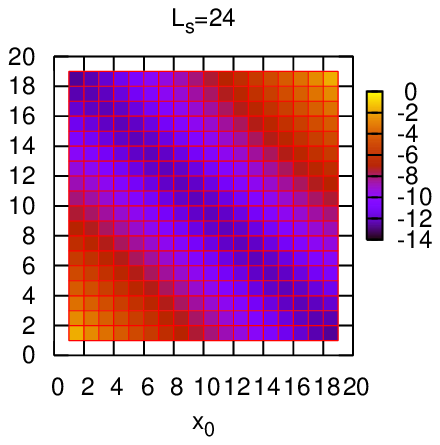}}\\
  \end{tabular}
\caption{$\log||\Delta^{(L_s)}(x_0,y_0)||_{\rm spin}$
for the zero spatial momentum with the parameters
$T=20$,
$m_{\rm f}=0$, $\theta=0$, $m_5=1$ and $c=1$.
\label{fig:GWbreaking}}
\end{center}
\end{figure}

Figure \ref{fig:GWbreaking} shows
the magnitude of the chiral symmetry breaking $\Delta^{(L_s)}$ in
eq.(\ref{eqn:GWbreaking})
with the zero spatial momentum configuration.
The norm in $||\Delta||_{\rm spin}$ is taken for the spinor index.
For $L_s=4$, we can see large chiral symmetry breaking
in diagonal elements
not only near boundaries but also in a bulk.
However, at increased $L_s$, the bulk breaking
gets smaller and then finally it disappears for $L_s=24$.
In the end, the remaining breaking is exponentially localized
near boundaries and we observe the expected behavior mentioned above.

\section{Concluding remarks and outlook}
By following the universality argument,
we constructed DWF in the SF setup.
And then we checked some fundamental
properties of the theory, namely, that
the lowest eigenmode is localized around the fifth dimensional boundaries
and the GW relation breaking for
the effective four dimensional operator has
the expected structure.
Although we do not show it in this paper,
we also checked the universality
at the one-loop level.
In total, our formulation works very well.

We are now ready to study NPR of $B_K$ for any flavors
with DWF by using the SF scheme.
Finally, we comment that
by using our formulation, one can carry out a search for
a conformal window.
Since chiral fermions have
benefits such as requiring no mass tuning to realize
the mass independent renormalization scheme
and no constraint on the number of flavors,
our formulation can play a crucial role
for solid quantitative studies.

\vspace{6mm}

S.T would like to thank Sinya Aoki, Yasumichi Aoki,
Norman Christ, Michael Endres,
Taku Izubuchi, Changhoan Kim
and the members of the RBC collaboration
for helpful discussions.
This work is supported by the U.S Department of Energy
under Grant No. DE-FG02-92ER40699.


\providecommand{\href}[2]{#2}\begingroup\raggedright\endgroup

\end{document}